\documentclass[%
 reprint,
showpacs,preprintnumbers,
nofootinbib,
 amsmath,amssymb,
 aps,
]{revtex4-1}

\usepackage{textcomp,booktabs}
\usepackage[usenames,dvipsnames]{color}
\usepackage{colortbl}
\definecolor{mygray}{gray}{.9}
\definecolor{mypink}{rgb}{.99,.91,.95}
\definecolor{mycyan}{cmyk}{.3,0,0,0}
\usepackage{makecell}
\usepackage{graphicx}
\usepackage{dcolumn}
\usepackage{bm}
\usepackage{graphics}
\usepackage{float}
\usepackage{tikz}
\usepackage{array}
\usepackage{verbatim}
\usepackage{pgflibraryarrows}
\usepackage{pgflibrarysnakes}
\usepackage{amsmath}
\usepackage{bm}
\usepackage{multirow}
\usepackage{array}
\usepackage{booktabs}

\allowdisplaybreaks[4]
\makeatletter

\newcommand{\Rmnum}[1]{\expandafter\@slowromancap\romannumeral #1@}
\makeatother
\newtheorem{theorem}{Theorem}

\newtheorem{proposition}{proposition}
\def\squareforqed{\hbox{\rlap{$\sqcap$}$\sqcup$}}
\def\qed{\ifmmode\squareforqed\else{\unskip\nobreak\hfil
\penalty50\hskip1em\null\nobreak\hfil\squareforqed
\parfillskip=0pt\finalhyphendemerits=0\endgraf}\fi}

\begin{document}


\title{Device-Independent Quantum Private Query Protocol without the Assumption of Perfect Detectors }
\author{Dan-Dan Li$^{1,2}$, Xiao-Hong Huang$^{1}$, Wei Huang$^{3}$, Fei Gao$^{4}$}
\affiliation{$^{1}$
Institute of Network Technology, Beijing University of Posts and Telecommunications, Beijing, 100876, China\\
$^{2}$ State Key Laboratory of Cryptology, P. O. Box 5159, Beijing, 100878, China\\
$^{3}$ Science and Technology on Communication Security Laboratory, Institute of Southwestern Communication, Chengdu 610041, China\\
$^{4}$ State Key Laboratory of Networking and Switching Technology,
Beijing University of Posts and Telecommunications, Beijing, 100876, China}
\begin{abstract}
The first device-independent quantum private query  protocol (MRT17) which is proposed by
Maitra \emph{et al.} [Phys. Rev. A 95, 042344 (2017)]  to
enhance the security through the certification of  the states  and measurements.
However, the MRT17 protocol works under  an assumption of perfect detectors, which increases difficulty
in the implementations.
Therefore, it is crucial to investigate what would affect the security of this protocol if
the detectors were imperfect.
Meanwhile, Maitra \emph{et al.} also pointed out that  this problem  remains open.
In this paper, we analyze the security of MRT17 protocol
when the detectors are imperfect and  then find that this  protocol is under attack in the aforementioned case.
Furthermore, we propose  device-independent QPQ protocol without the assumption of perfect detectors.
Compared with MRT17 protocol, our  protocol is more practical without relaxing the security in the device-independent framework.

\end{abstract}

\maketitle


\section{Introduction}
Private information retrieval (PIR) \cite{1} deals with the problem that an user (Alice)
knows the address of an item from the database  with $N$ items which is held by Bob and queries it secretly.
In a PIR protocol,  Alice gets  correctly the item   that she queried, whereas Bob does
not know which item Alice has queried (i.e., the perfect user privacy).
Furthermore,
a symmetrically private information retrieval (SPIR) \cite{2} has
  one more security requirement that Alice cannot get other items from the database except what she queried (i.e., the perfect database security).
However, the task of SPIR cannot be realized ideally  even in quantum cryptography \cite{3}.

As a quantum protocol for dealing with SPIR problems, quantum private query (QPQ)
relaxes the security requirements to some extent:
\romannumeral 1) Alice has nonzero probability to discover Bob's attack if he attempts to learn the address of Alice's queried item,
which is referred to as cheat sensitivity.  \romannumeral 2) Alice can gain  a few more  items than the perfect requirement where she only obtains the queried item.

In 2008, Giovannetti \emph{et al.} \cite{4}  proposed the first cheat-sensitive QPQ protocol (GLM08),
where the database was represented by a unitary operation (i.e., oracle operation) and it was performed on the query/test states at random, which
were prepared by Alice. The query states were to obtain the retrieved item and the test states were to check potential attack from Bob.
This protocol reduces exponentially the communication and computation complexity.
Furthermore, the security of GLM08 protocol has been analyzed strictly \cite{5} and a proof-of-principle experiment   has been implemented \cite{6}.
 Olejnik  improved GLM08 protocol such that communication complexity was reduced further (O11)\cite{7}.
These  two protocols exhibit significant advantage in theory; but for large database dimension,  these protocols are not practically implementable.
 To solve this problem,  Jacobi \emph{et al.} \cite{8} proposed a QPQ protocol (J11) based on   SARG04 quantum key distribution (QKD) protocol \cite{9}, where this kind of protocols were called as QKD-based QPQ protocols.
 Compared with GLM08  and O11 protocols, J11 protocol can tolerate losses
  and be easily implemented to a  database with large dimension.
Subsequently,
 many QKD-based QPQ protocols \cite{10,11,12} are proposed and their security   either better users privacy or better database security even the perfect database security is studied \cite{133,13,14,144}.
  The development of QPQ protocols can be referred to \cite{123}.

\begin{figure}\centering
\includegraphics[width=8.6cm]{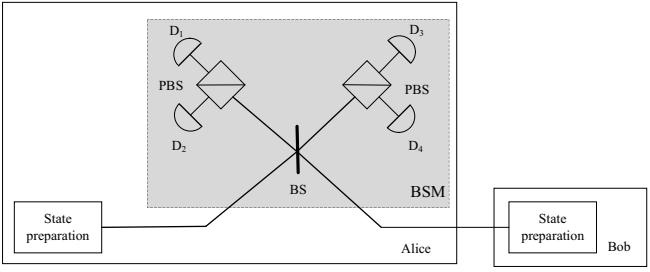}
\caption{The framework of MDI QPQ.  MDI indicates that measurement devices  need not be characterized any more.
The grey dotted box represents that Alice need not trust the Bell-state-measurement (BSM) device.  BS and PBS represent the 50:50 beam splitter and polarization beam splitter, respectively.}
\label{fig0}
\end{figure}

\begin{figure}
\centering
\includegraphics[width=8.6cm]{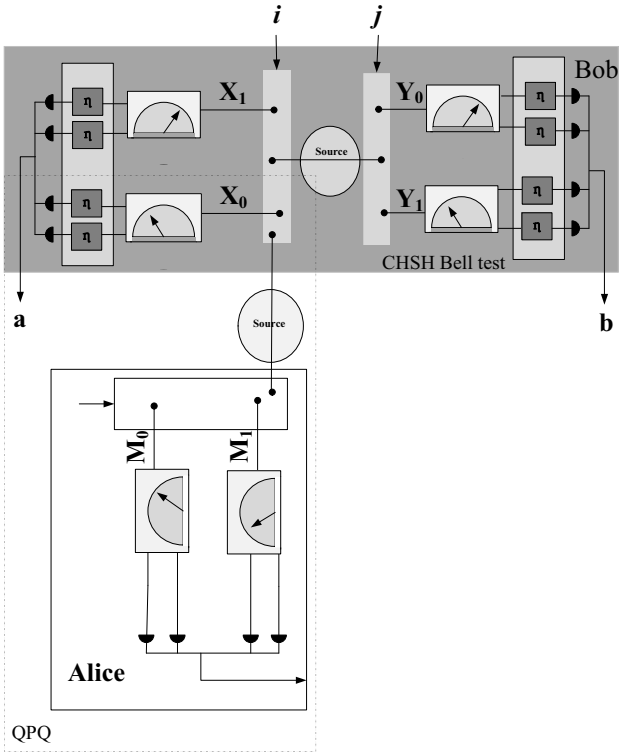}
\vspace{1em}
\caption{The framework of DI-QPQ. The grey box represents that Bob need not trust the device of states and measurements
but certify the states and measurements via local CHSH Bell game.
When the observed value violates the critical value of CHSH Bell game, QPQ part can be performed. Note that $\eta$ represents detection efficiency
and  assume  $\eta=1$ in MRT17 protocol.}
\label{fig2}
\end{figure}

 In the traditional QPQ protocol, Alice and Bob  are required to trust their devices. If the states shared between them are not in the predetermined form, then Alice can always utilize some strategies which help
 her to elicit a few more items than what is suggested by the protocol. Thus, it is necessary for Bob  to certify the states and measurements.
With the advent of device-independent idea, the related cryptographic protocols do not rely on any assumptions about the states and measurements in their protocols. Zhao \emph{et al.}  \cite{15} designed measurement device independence (MDI) QPQ protocol to
 certify the   measurements (see Fig. \ref{fig0}). Later,  Maitra, Paul and Roy proposed a device-independent  (DI) QPQ protocol (MPR17) \cite{16} to
 certify both states and measurements before proceeding to QPQ part (see Fig. \ref{fig2}).

Obviously, MRT17 protocol enhances the security since   the states and measurements need not
 be trusted but can be certified whether to be in a predetermined form.
While, this protocol works under an assumption of   perfect detectors (i.e., detectors with unit efficiency),
 which is hard to implement  experimentally. If the assumption is relaxed, how will it affect the security of MRT17 protocol?

In this paper, we solve this problem and propose a DI-QPQ protocol without the assumption of perfect detectors.
Firstly, we analyze security threat for MRT17 protocol under the case that  the assumption is relaxed and the results
show that this protocol is under attack in the above case.
Secondly, we propose DI-QPQ protocols without the assumption of perfect detectors.
Compared with MRT17 protocol, our protocol not only  maintains the security in the DI framework but also is toward practical.

The remaining of the paper is as follows. The review of DI-QPQ protocol with an assumption of perfect detectors is given in Sect. \ref{s2}.
DI-QPQ protocol without the assumption of perfect detectors is proposed and the security analysis is given in Sect. \ref{s3}.
The conclusion is summarized in the last section.

\section{Review  DI-QPQ protocol with the assumption of perfect detectors }
\label{s2}
The MRT17 protocol \cite{16} is the first QPQ protocol in the DI framework. Our protocol is built on MRT 17 protocol. In this section, we revisit  MRT17 (DI-QPQ) protocol \cite{16}.

 In the DI framework,  the assumptions about  states  and measurements  need not be made except for the basic ones.
In the MRT17 protocol, the trustworthiness of Bob's devices  can be removed, and Bob utilizes a statistical method (known as  CHSH Bell game) \cite{17}
to certify whether the shared states and measurements are in the predetermined  forms between them.

Before introducing the details of MRT17 protocol, we give the relation between CHSH Bell test and CHSH Bell game.

In the standard CHSH Bell test,
\begin{enumerate}
  \item $x_{i}\in\{0,1\}$  and $y_{i}\in\{0,1\}$ are chosen uniformly at random.
   $x_{i}=0 (1)$ represents the measuring observables $\sigma_z (\sigma_x)$; $y_{i}=0 (1)$ represents  the measuring  observables $\frac{\sigma_{x}+\sigma_{z}}{\sqrt{2}} (\frac{\sigma_{z}-\sigma_{x}}{\sqrt{2}})$. The first and second particle of the entangled state can be measured, respectively.
 When the measurement result is $|0\rangle$ or $|+\rangle$ ($|1\rangle$ or $|-\rangle$), then $a_{i}=0 (1)$.
Similarly,  the definition of $b_{i}$ has been given.
  \item  CHSH Bell correlation function is described as
\begin{equation*}
\begin{split}
\hat{I}_{CHSH}&=E(X_{0}Y_{0})+E(X_{0}Y_{1})
+E(X_{1}Y_{0})-E(X_{1}Y_{1}),
\end{split}
\end{equation*}
where
\begin{equation*}
E(X_{0}Y_{0})
=\sum_{a_{i}, b_{i}}(-1)^{a_{i}+ b_{i}}p(a_{i}, b_{i}|x_{i}=0, y_{i}=0),
\end{equation*}
and others have similar definitions.
\end{enumerate}
 Now, if the shared state is $\frac{|00\rangle+|11\rangle}{\sqrt{2}}$, then
$\hat I_{CHSH}=2\sqrt{2},$
which corresponds to the maximal violation in quantum theory.

In a general CHSH Bell test, the shared state is not the maximally  entangled state  but general entangled state and the measuring observables are $\{|\psi\rangle, |\psi^{\bot}\rangle\}$ for different four angles,
where
\begin{equation}
\begin{split}
&|\psi\rangle=\cos\frac{\psi}{2}|0\rangle+\sin\frac{\psi}{2}|1\rangle, \\
&|\psi^{\bot}\rangle=\sin\frac{\psi}{2}|0\rangle-\cos\frac{\psi}{2}|1\rangle.\\
\end{split}
\end{equation}
thus $\hat I_{CHSH}\in (2, 2\sqrt{2}]$ in quantum theory.

The CHSH Bell test is considered as  a nonlocal game. Two players ($\mathfrak{P}_{1}, \mathfrak{P}_{2}$)
 are viewed as cooperating with each other.
A referee runs the game, and all communication is between the players and referee, while no communication
directly between the players is permitted. The referee selects randomly $x_{i}y_{i}\in\{00, 01, 10, 11\}$.
Then, each player must answer a single bit ($a_{i}$ for $\mathfrak{P}_{1}$, $b_{i}$ for $\mathfrak{P}_{2}$).
They win if $a_{i}+b_{i}=x_{i}\wedge y_{i}$.
 CHSH correlation function is characterized in the CHSH Bell test, while the average probability of success
is described in the CHSH Bell game. The average probability of success$=\frac{1}{2}+\frac{\hat I_{CHSH}}{8},$ \cite{}.

Next, we review the MRT17 protocol.

Firstly,  the basic assumptions in the MRT17 protocol are listed \cite{16}:
\begin{description}
  \item[A]  The additional information in  Alice and Bob's laboratories is not leaked.
  \item[B] Each use of device is independent of the previous uses.
  \item[C] All the detectors of Bob are perfect, i.e., unit efficiency.
\end{description}

The details of MRT17 protocol are introduced \cite{16}.
\begin{enumerate}
\item Bob starts with $N$ entangled states which may be  prepared by Alice.

\item Bob chooses some entangled states at random, which forms $\Gamma_{CHSH}$ for  CHSH Bell test. $\Gamma_{CHSH}$ contains $\lceil\gamma N\rceil$ entangled states, where $0 <\gamma< 1$. The remained entangled states constitute $\Gamma_{QPQ}$ for QPQ
protocol which contains $\lfloor(1-\gamma)N\rfloor$ entangled states.

\item For CHSH Bell game, $i\in\{1,..., \lceil\gamma N\rceil\}$,

\begin{enumerate}
  \item  Bob chooses $x_{i}\in\{0,1\}$  and $y_{i}\in\{0,1\}$ uniformly at random.
  \item If
$x_{i}=0 (1)$, he measures the first particle of the entangled state in the basis $\{|0\rangle, |1\rangle\} (\{|+\rangle, |-\rangle\})$,
then denote $a_{i}=0 (1)$ when the measurement result is $|0\rangle$ or $|+\rangle$ ($|1\rangle$ or $|-\rangle$).
In the same manner, if
$y_{i}=0 (1)$, he measures the second particle of the entangled state in the basis $\{|\psi_{1}\rangle,  |\psi_{1}^{\bot}\rangle\} (\{|\psi_{2}\rangle, |\psi_{2}^{\bot}\rangle\})$,
then denote $b_{i}=0 (1)$ when the measurement result is $|\psi_{1}\rangle$ or $|\psi_{2}\rangle$ ($|\psi_{1}^{\bot}\rangle$ or $|\psi_{2}^{\bot}\rangle$), where
\begin{equation}
\begin{split}
&|\psi_{l}\rangle=\cos\frac{\psi_{l}}{2}|0\rangle+\sin\frac{\psi_{l}}{2}|1\rangle, \\
&|\psi_{l}^{\bot}\rangle=\sin\frac{\psi_{l}}{2}|0\rangle-\cos\frac{\psi_{l}}{2}|1\rangle\\
\end{split}
\end{equation}
 for $l\in\{1, 2\}$, $\psi_{1}\in (0, \frac{\pi}{2})$, $\psi_{2}\in (\frac{\pi}{2},\pi)$ and $\psi_{1}+\psi_{2}=\pi$.





  \item  \emph{Statistical method 1:} define $\mathcal{Y}_{i}$ be the observed result of the $i$th game for players, i.e.,
\begin{equation}
\mathcal{Y}_{i}=
\begin{cases}
&1, \qquad\text{if}\quad a_{i}+b_{i}=x_{i}\wedge y_{i},\\
&0, \qquad\text{otherwise},
\end{cases}
\label{t1}
\end{equation}
and let $\mathcal{Y}$ be the average observed probability of success in the CHSH Bell game,
\begin{equation}
\mathcal{Y}=\frac{1}{\lceil\gamma N\rceil}\sum_{i}\mathcal{Y}_{i}.
\label{t2}
\end{equation}
If
\begin{equation}
\mathcal{ Y} <\bar{\mathcal{ Y}},
\label{tt2}
\end{equation}
then Bob aborts the protocol, where
$\bar{\mathcal{ Y}}$ represents the  average probability of success,
 \begin{equation}
\begin{split}
\bar{\mathcal{ Y}} =&\frac{1}{8}[\sin\theta(\sin\psi_{1}+\sin\psi_{2})+\cos\psi_{1}\\
 &-\cos\psi_{2}]+\frac{1}{2}.
  \end{split}
  \label{e2}
  \end{equation}
Otherwise, he  proceeds to the following QPQ protocol \cite{11}.
\end{enumerate}

\item  For QPQ part, the condition must be satisfied that   the local CHSH Bell game at Bob's end violates the above relation (\ref{tt2}), thus the states  shared between Alice and Bob are certified to be in their predetermined form, i.e.,
\begin{equation}
  \Psi_{BA}=\frac{1}{\sqrt{2}}(|0\rangle_{B}|\phi_{0}\rangle_{A}+|1\rangle_{B}|\phi_{1}\rangle_{A}),
  \label{e1}
  \end{equation}
  where
      \begin{align}
      &|\phi_{0}\rangle_{A}=\cos\frac{\theta}{2}|0\rangle+\sin\frac{\theta}{2}|1\rangle,\\
      &|\phi_{1}\rangle_{A}=\cos\frac{\theta}{2}|0\rangle-\sin\frac{\theta}{2}|1\rangle,
      \end{align}
      for $\theta\in (0, \frac{\pi}{2})$.

Next, Bob uses the remaining $\lfloor(1-\gamma)N\rfloor$ entangled pairs to  proceed to QPQ steps.

\begin{enumerate}
  \item Bob  sends a particle $A$ in each certified entangled pair  (\ref{e1}) to Alice.

  \item Alice announces whether  she has successfully received the particle or not.
  For un-received particle of Alice, Bob discards the
  corresponding particle.

  \item Then, Bob measures his particle in the basis $\{|0\rangle, |1\rangle\}$, Alice measures
   the corresponding particle in the basis $\{|\phi_{0}\rangle, |\phi_{0}^{\perp}\rangle\}$ or $\{|\phi_{1}\rangle, |\phi_{1}^{\perp}\rangle\}$. If
  Alice's measurement outcome is $|\phi_{0}^{\perp}\rangle$ ( $|\phi_{1}^{\perp}\rangle$),
  she can conclude that the raw key bit at Bob's end is 1 (0),  and then the success probability that Alice gains a bit is showed in Table \ref{ta2}.
\begin{table}
\centering
\caption {The success probability that Alice obtains a  key.}
\begin{tabular}{c|cc>{\columncolor{blue!25}}cc>{\columncolor{blue!25}}c}
\hline
\hline
\specialrule{0em}{3pt}{3pt}
classical coding &&  \ 0\ & 1 & 1 & 0\\
\specialrule{0em}{3pt}{3pt}
&&  \ $|\phi_{0}\rangle_{A}$\ & $|\phi_{0}^\perp\rangle_{A}$ & $|\phi_{1}\rangle_{A}$ & $|\phi_{1}^\perp\rangle_{A}$\\
\hline
\specialrule{0em}{3pt}{3pt}
\quad 0&$\quad |0\rangle_{B}$  & $\frac{1}{2}$ &0 & $\frac{1}{2}\cos^{2}\theta$ & $\frac{1}{2}\sin^{2}\theta$\\
\specialrule{0em}{3pt}{3pt}
\quad 1&$\quad |1\rangle_{B}$  & $\frac{1}{2}\cos^{2}\theta$ & $\frac{1}{2}\sin^{2}\theta$ & $\frac{1}{2}$  &0\\
\hline
\hline
\end{tabular}
\label{ta2}
\end{table}

\item Alice's measurements  yield  conclusive results and
inconclusive ones. Both conclusive and inconclusive results are stored. Alice and Bob now share a
raw key string (i.e., $K$).
Bob knows the whole raw key string, whereas Alice generally knows the part of $K$, i,e., $\frac{\sin^2\theta}{2}|K|$,
where $|K|$ represents the number of $K$.

\item Alice and Bob postprocess  $K$ so that Alice's known key bits of  $K$ are reduced
to 1 bit.

\item Bob encrypts his database so that  Alice only obtains the item that she queried.
\end{enumerate}
\end{enumerate}

Different from  traditional QPQ protocol, MRT17 protocol can certify whether the states and measurements to
be in their predetermined forms.
Thus, MRT17 protocol has an advantage of improving the security.

\section{DI-QPQ protocol without the assumption of perfect detectors}
\label{s3}
In this section, we discuss the security of MRT17 protocol when the assumption is relaxed.
Then, we propose DI-QPQ protocol without the assumption of perfect detectors.

In general, the security analysis of QKD introduces an outside adversary (Eve) and investigates the effect of Eve's attack,
then rules out it.
Different from QKD,  each of the two  parties may be an attacker for the counterpart in a QPQ protocol.
Two cases are included  in the security analysis: \romannumeral 1) Alice tries to extract  more information about the raw keys,
or \romannumeral 2) Bob tries his best to learn the address of item that Alice queries.

\subsection{The attack against MRT17 protocol if the detectors are imperfect}
By considering a general attack scenario that Alice, as an adversary, tries to elicit more information from the raw keys,
we discuss the security of MRT17 protocol with the assumption of perfect detectors via two statistical methods from CHSH Bell
game/test, respectively. Furthermore, when the assumption is relaxed, we show that MRT17 protocol is under attack.

 Suppose that Alice has ability to prepare the biased states
 such that  the state is in the following form
\begin{equation}
\alpha|0\rangle_{B}|\phi_{0}\rangle_{A}+\beta|1\rangle_{B}|\phi_{1}\rangle_{A},
\label{e6}
\end{equation}
where
\begin{equation}
\begin{split}
&\alpha=\sqrt{\frac{1}{2}+\epsilon}, \qquad
\beta=\sqrt{\frac{1}{2}-\epsilon},\\
 &\epsilon\in (-\frac{1}{2}, 0)\cup (0, \frac{1}{2}),
 \end{split}
 \end{equation}
the difference between them lies in $\alpha, \beta$ compared with the predetermined state (\ref{e1}).

Firstly, we show that Alice's attack does not work for MRT17 protocol with the assumption of perfect detector no
matter which statistical method is from CHSH Bell game/test.

\noindent(1) the shared state in MRT17 protocol is certified in the way of nonlocal game.

After local CHSH Bell game, Bob  gets

\begin{equation}
\begin{split}
\mathcal{Y}&=\frac{1}{8}\sin\theta(\sin\psi_{1}+\sin\psi_{2})+\frac{\sqrt{\frac{1}{4}-\epsilon^2}}{4}(\cos\psi_{1}-\cos\psi_{2})\\
&
\quad +\frac{1}{2},
\end{split}
\label{e3}
\end{equation}
this value (\ref{e3}) cannot violate the condition (\ref{tt2}) for an arbitrary value of $\epsilon$
(i.e., $\in (-\frac{1}{2}, 0)\cup (0, \frac{1}{2})$),  and the proof is given  Appendix A. According to the rules of MRT17 protocol,   this protocol can be terminated.
Obviously, Alice's  attack does not work for  MRT17 protocol
with the prefect detectors.

\noindent(2) the shared state in MRT17 protocol is certified in the way of CHSH Bell test.
Statistical method of CHSH Bell test is in the following:

 instead of steps (\ref{t1})-(\ref{t2}),
\emph{Statistical method 2}:
denote $I_{i}$ as the observed result in the $i$th experiment.
\begin{equation}
\begin{split}
  &I_{i}\\
&=\sum_{a_{i},b_{i},x_{i},y_{i}}(-1)^{a_{i}\oplus b_{i}\oplus x_{i}y_{i}}\frac{\chi(a_{i}=a,b_{i}=b,x_{i}=x,y_{i}=y)}{p(x_{i},y_{i})},\\
  \end{split}
  \label{e0}
  \end{equation}
where
\begin{equation}
\chi(x)=
\left\{
  \begin{array}{ll}
    1, & \hbox{$x$ is observed;} \\
    0, & \hbox{otherwise;}
  \end{array}
\right.
\end{equation}

$I_{CHSH}$ represents the observed value of CHSH Bell correlation function, i.e.,
  \begin{equation}
  I_{CHSH}=\frac{1}{\lceil\gamma N\rceil}\sum_{i=1}^{\lceil\gamma N\rceil}I_{i}.
  \label{e30}
  \end{equation}
When $N\rightarrow \infty$, $\hat I_{CHSH}$ can be represented as 
\begin{equation}
\begin{split}
\hat{I}_{CHSH}=&\frac{1}{\lceil\gamma N\rceil}\sum_{i=1}^{\lceil\gamma N\rceil}
\sum_{a_{i},b_{i},x_{i},y_{i}}(-1)^{a_{i}+b_{i}+x_{i}y_{i}}\frac{\chi(a_{i}=a,}{}\\
&\quad\frac{b_{i}=b,x_{i}=x,y_{i}=y)}{p(x_{i},y_{i})}\\
=&\frac{1}{p(x_{i},y_{i})}\sum_{i=1}^{\lceil\gamma N\rceil}\sum_{a_{i},b_{i},x_{i},y_{i}}(-1)^{a_{i}+b_{i}+x_{i}y_{i}}
\frac{\chi(a_{i}=a,}{}\\
&\quad\frac{b_{i}=b,x_{i}=x,y_{i}=y)}{\lceil\gamma N\rceil}\\
=&\frac{1}{p(x,y)}\sum_{a,b,x,y}(-1)^{a+b+xy}p(a,b,x,y)\\
=&\sum_{a,b,x,y}(-1)^{a+b+xy}p(a,b|x,y)\\
=&\sum_{a,b}(-1)^{a+b}[p(a,b|X_{0},Y_{0})+p(a,b|X_{0},Y_{1})+\\
&p(a,b|X_{1},Y_{0})-p(a,b|X_{1},Y_{1})\\
=&E(X_{0}Y_{0})+E(X_{0}Y_{1})+E(X_{1}Y_{0})-E(X_{1}Y_{1}).
\end{split}
\end{equation}

If
\begin{equation}
I_{CHSH}<\hat{I}_{CHSH},
\label{tt3}
\end{equation}

where
\begin{equation}
\hat{I}_{CHSH}=\sin\theta(\sin\psi_{1}+\sin\psi_{2})+\cos{\psi_{1}-\cos{\psi_{2}}},
\label{e5}
\end{equation}
 Bob aborts the protocol. Otherwise, Bob proceeds to the QPQ part. Note that
the deduction of (\ref{e5}) can be referred to  Appendix B.

Hence, based on the state (\ref{e6}) and \emph{statistical method} 2, Bob gets
\begin{equation}
\begin{split}
I_{CHSH}&=\sin\theta(\sin\psi_{1}+\sin\psi_{2})
+2\sqrt{\frac{1}{4}-\epsilon^2}(\cos\psi_{1}\\
&\quad-\cos\psi_{2}).
\end{split}
\label{e4}
\end{equation}

This value cannot violate the condition (\ref{tt3}), the proof can be given  in Appendix C.

To summarize, these  two statistical methods can detect Alice's attack that the states are
not in the predetermined forms.
 These two statistical methods are equivalent.

Next, when the assumption is relaxed, we show that MRT17 protocol  can be attacked successfully
via Alice's strategy that uses biased state (\ref{e6}).

We make use of
statistical method 2 to catch the influence of the imperfect detectors.
When the detection efficiency of Bob's detectors
 is not limited to be unit,  denoted as  $\eta$, i.e.,
\begin{equation}
\eta=\min_{j,k}\{p(\Lambda_{X_{j}}), p(\Lambda_{Y_{k}})\}
\end{equation}
where $\Lambda_{X_{j}}(\Lambda_{Y_{k}})$ represents the subensemble of  nonempty outputs (i.e,  $a (b)\in\{0, 1\} $) when choosing
measurement $X_{j} (Y_{k})$.

When the detectors are imperfect, the value of $I_{CHSH}$ with the biased state (\ref{e6}) is
\begin{equation}
\begin{split}
I_{CHSH}&=\frac{8-2[\sin\theta(\sin\psi_{1}+\sin\psi_{2})+2\sqrt{\frac{1}{4}-\epsilon^{2}}}{\eta}\\
&\frac{
(\cos\psi_{1}-\cos\psi_{2})]}{}+3[\sin\theta(\sin\psi_{1}+\sin\psi_{2})+\\
&2\sqrt{\frac{1}{4}-\epsilon^{2}}(\cos\psi_{1}-\cos\psi_{2})]-8.
\end{split}
\label{e21}
\end{equation}

When $\eta$ and $\epsilon$ satisfy the following cases:

\noindent Case 1:
\begin{equation}
\begin{split}
&\frac{2}{1+\sqrt{2}}<\eta<\frac{8-2A}{8-2A+B}<1,\\
& \epsilon \in (-\frac{1}{2}, 0)\cup (0, \frac{1}{2});
\end{split}
\end{equation}

\noindent Case 2:
\begin{equation}
\begin{split}
&\frac{2}{1+\sqrt{2}}<\frac{8-2A}{8-2A+B}<\eta<\frac{8-2A-2B}{8-2A-2B},\\
&\epsilon \in (-\frac{\sqrt{(3\eta-2)^2B^2-C^2}}{2(3\eta-2)B}, 0)\cup (0, \frac{\sqrt{(3\eta-2)^2B^2-C^2}}{2(3\eta-2)B}),
\end{split}
\end{equation}

where
\begin{align}
&A=\sin\theta(\sin\psi_{1}+\sin\psi_{2}),\\
&B=\cos\psi_{1}-\cos\psi_{2},\\
&C=(8-2A+B)\eta-8+2A.
\end{align}

Eq.(\ref{e21})  violates the relation (\ref{tt3})(the proof  is given in Appendix D),   thus Bob proceeds to QPQ part.
In the above cases, Alice successfully cheats  Bob
to make Bob believe that the states in his hand are in predetermined form which actually are not.
In this way,  the success probability of Alice's guessing  keys from the raw keys is $(\frac{1}{2}+2\epsilon^2)\sin^2\theta$.
Compared with $p=\frac{\sin^2\theta}{2}$, Alice successfully gets more information from Bob's database without being caught. Therefore,
when the detectors are not perfect, MRT17 protocol is insecure.

\subsection{DI-QPQ protocol without the assumption of perfect detectors}

According to the part A, we find that MRT17 protocol suffers an attack when the detectors are imperfect.
How to design a DI-QPQ protocol without the assumption of perfect detectors becomes crucial.
We rule out the influence of detector inefficiency  and then propose  DI-QPQ protocol without the assumption of perfect detectors.
Our assumptions now are relaxed to the first two  ones of MRT17 protocol.
That is,
\begin{description}
  \item[A] the additional information in  Alice and Bob's laboratories is no leaked.
  \item[B] each use of device is independent of the previous uses.
\end{description}

Denote the detection efficiency of Bob's detectors as $\eta$. The procedures of DI-QPQ protocol without the assumption of perfect detectors are as follows:
\begin{enumerate}
\item Bob starts with $N$ entangled states which may be  prepared by Alice.

\item Bob chooses some entangled states at random, which forms $\Gamma_{CHSH}$ for  CHSH Bell test. $\Gamma_{CHSH}$ contains $\lceil\gamma N\rceil$ entangled states, where $0 <\gamma< 1$. The remained entangled states constitute $\Gamma_{QPQ}$ for QPQ
protocol which contains $\lfloor(1-\gamma)N\rfloor$ entangled states.

\item For CHSH Bell test, $i\in\{1,..., \lceil\gamma N\rceil\}$,

\begin{enumerate}
  \item  Bob chooses $x_{i}\in\{0,1\}$  and $y_{i}\in\{0,1\}$ uniformly at random.
  \item If
$x_{i}=0 (1)$, he measures the first particle of the entangled state in the basis $\{|0\rangle, |1\rangle\} (\{|+\rangle, |-\rangle\})$,
then denotes $a_{i}=0 (1)$ when the measurement result is $|0\rangle$ or $|+\rangle$ ($|1\rangle$ or $|-\rangle$).
In the same manner, if
$y_{i}=0 (1)$, he measures the second particle of the entangled state in the basis $\{|\psi_{1}\rangle,  |\psi_{1}^{\bot}\rangle\} (\{|\psi_{2}\rangle, |\psi_{2}^{\bot}\rangle\})$,
then denotes $b_{i}=0 (1)$ when the measurement result is $|\psi_{1}\rangle$ or $|\psi_{2}\rangle$ ($|\psi_{1}^{\bot}\rangle$ or $|\psi_{2}^{\bot}\rangle$), where
\begin{equation}
\begin{split}
&|\psi_{l}\rangle=\cos\frac{\psi_{l}}{2}|0\rangle+\sin\frac{\psi_{l}}{2}|1\rangle, \\
&|\psi_{l}^{\bot}\rangle=\sin\frac{\psi_{l}}{2}|0\rangle-\cos\frac{\psi_{l}}{2}|1\rangle\\
\end{split}
\end{equation}
 for $l\in\{1, 2\}$, $\psi_{1}\in (0, \frac{\pi}{2})$, $\psi_{2}\in (\frac{\pi}{2},\pi)$ and $\psi_{1}+\psi_{2}=\pi$.

  \item \emph{Statistical method 2}:
\begin{equation}
\begin{split}
  &I_{i}\\
  &=\sum_{a_{i},b_{i},x_{i},y_{i}}(-1)^{a_{i}+b_{i}+x_{i}y_{i}}\frac{\chi(a_{i}=a,b_{i}=b,x_{i}=x,y_{i}=y)}{p(x_{i},y_{i})},\\
  \end{split}
  \end{equation}
where $I_{i}$ represents the observed result in the $i$th experiment.
  \begin{equation}
  I_{CHSH}=\frac{1}{\lceil\gamma N\rceil}\sum_{i=1}^{\lceil\gamma N\rceil}I_{i},
  \label{t3}
  \end{equation}
$I_{CHSH}$ represents the observed value of CHSH Bell correlation function.
  \item If
  \begin{equation}
  I_{CHSH}<\bar{I}_{CHSH},
  \label{ee1}
  \end{equation}
  where
  \begin{equation}
  \begin{split}
  \bar{I}_{CHSH}&= -8+3[\sin\theta(\sin\psi_{1}+\sin\psi_{2})+\cos\psi_{1}-\cos\psi_{2}]\\
      &\quad+\frac{8-2[\sin\theta(\sin\psi_{1}+\sin\psi_{2})+\cos\psi_{1}-\cos\psi_{2}]}{\eta},
  \end{split}
  \label{s33}
  \end{equation}
$\bar{I}_{CHSH}$ represents the value of CHSH Bell correlation function.
\item  For QPQ part, the condition must be satisfied that   the local CHSH Bell test at Bob's end violates the above relation (\ref{ee1}), thus the states  shared between Alice and Bob are certified to be in their predetermined form, i.e.,
\begin{equation}
  \Psi_{BA}=\frac{1}{\sqrt{2}}(|0\rangle_{B}|\phi_{0}\rangle_{A}+|1\rangle_{B}|\phi_{1}\rangle_{A}),
  \end{equation}
  where
      \begin{align}
      &|\phi_{0}\rangle_{A}=\cos\frac{\theta}{2}|0\rangle+\sin\frac{\theta}{2}|1\rangle,\\
      &|\phi_{1}\rangle_{A}=\cos\frac{\theta}{2}|0\rangle-\sin\frac{\theta}{2}|1\rangle,
      \end{align}
      for $\theta\in (0, \frac{\pi}{2})$.

Next, Bob uses the remaining $\lfloor(1-\gamma)N\rfloor$ entangled pairs to  proceed to QPQ steps which are the same as that of MRT17 protocol.
\end{enumerate}
\end{enumerate}
Note that the deduction of $\bar{I}_{CHSH}$ can be  referred to Appendix E

\begin{figure}[h]\centering
\includegraphics[scale=0.7]{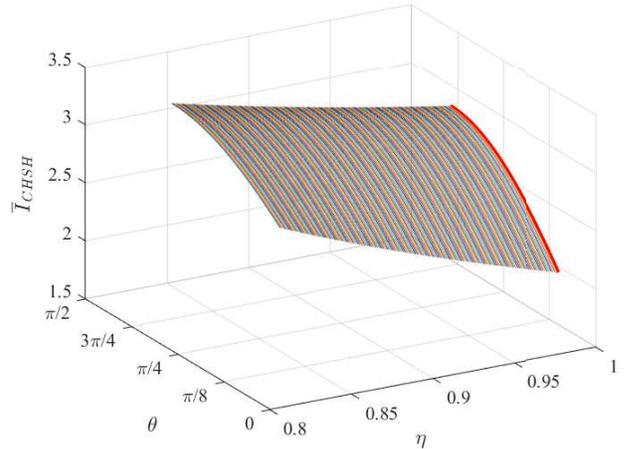}
\caption{The relation among $\theta$, $\eta$ and the value of $\bar{I}_{CHSH}$ when $\psi_{1}=\pi/4, \psi_{2}=3\pi/4$. The red line represents the result of MRT17 protocol with assumption of perfect detectors.}
\label{fig3}
\end{figure}

 In the following, we show the relation among $\theta$, detection efficiency ($\eta$) and
 the value of $\bar{I}_{CHSH}$ when $\psi_{1}=\pi/4, \psi_{2}=3\pi/4$ (see Fig. {\ref{fig3}}).

From Fig. \ref{fig3}, the value of $\bar{I}_{CHSH}$  rules out the effect of detector inefficiency.
 When $I_{CHSH}=\bar{I}_{CHSH}$,   the shared state is  predetermined one; otherwise it is not.
 In particular, we show the relations between $\bar{I}_{CHSH}$ and $\theta$ when fixing the values of detection
efficiency, i.e., $\eta=1, \ 0.83$, respectively.
\begin{figure}[h]\centering
\includegraphics[scale=0.7]{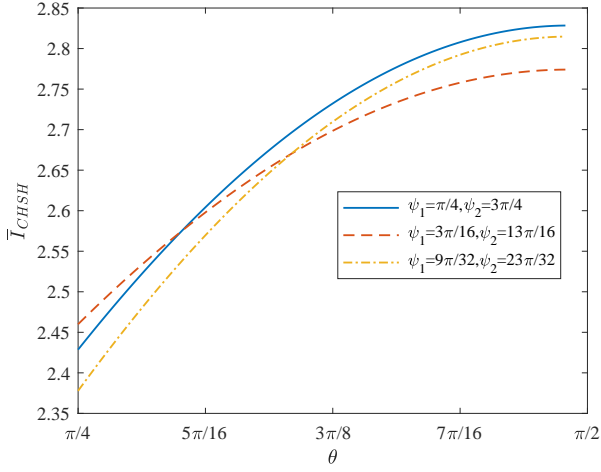}
\caption{The relation between the value of $\bar{I}_{CHSH}$ and $\theta$ conditioned that $\eta=1$. The three lines represent these relations
of different values of  $(\psi_{1}, \psi_{2})=(\pi/4, 3\pi/4),(3\pi/16, 13\pi/16)$ and $(9\pi/32, 23\pi/32)$, respectively.}
\label{fig4}
\end{figure}

\begin{figure}[h]\centering
\includegraphics[scale=0.7]{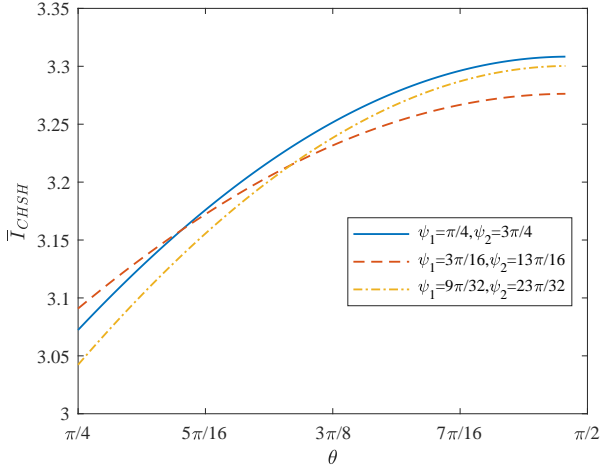}
\caption{The relation between the value of $\bar{I}_{CHSH}$ and $\theta$ conditioned that $\eta=0.83$.
The three lines represent these relations when $(\psi_{1}, \psi_{2})=(\pi/4, 3\pi/4),(3\pi/16, 13\pi/16)$ and $(9\pi/32, 23\pi/32)$, respectively.}
\label{fig5}
\end{figure}

\subsection{Security Analysis}
In this section, we analyze the security of our protocol.

\begin{theorem}
Suppose that there are $N$ pairs of entangled states in DI-QPQ protocol without the assumption of perfect detectors,
these entangled states are divided randomly into two parts: $\Gamma_{CHSH}\subset\{1, \cdots, N\}$ of size $\lceil\gamma N\rceil$ with $0<\gamma<1$ and $\Gamma_{QPQ}=\{1,\cdots, N\} \backslash \Gamma_{CHSH}$.
  If
\begin{equation}
I_{CHSH}=\bar{I}_{CHSH}-\xi
\label{s1}
\end{equation}
 is satisfied in the local CHSH Bell test, and then the  relation (\ref{s1}) is still
satisfied  in the QPQ part with a negligible statistical deviation $v$,
where
$I_{CHSH}$ represents the observed value of CHSH Bell test, and $\bar{I}_{CHSH}$ represents the  value of CHSH Bell test in
predetermined states and measurements in our protocol.
\begin{equation}
\begin{split}
\xi=\sqrt{\frac{1}{2\lceil\gamma N\rceil}\ln \frac{1}{\varepsilon_{CHSH}}},\\
v=\sqrt{\frac{(N-\lceil\gamma N\rceil+1)\lfloor(1-\gamma)N\rfloor^2
\ln\frac{1}{\varepsilon_{QPQ}}}{2\lceil\gamma N\rceil N^3}},
\end{split}
\end{equation}
 $\varepsilon_{CHSH}$ and $\varepsilon_{QPQ}$ are negligible small values.
\end{theorem}
\label{pro1}

\emph{proof}: Different from the proof of MRT17 protocol, $I_{i}$ can be rewritten as
\begin{equation}
I_{i}=
\begin{cases}
\frac{(-1)^{a+b+xy}}{p(x,y)},& \text{if}\ (a_{i}, b_{i}, x_{i}, y_{i})=(a,b,x,y) \ \text{is observed;}\\
0, & \text{otherwise.}\\
\end{cases} \\
\end{equation}
Then, $I_{CHSH}=\frac{1}{\lceil\gamma N\rceil}\sum_{i\in\Gamma_{CHSH}}I_{i}$ is
called as the observed average value. The expected value of $I_{CHSH}$ is $\bar{I}_{CHSH}$.

Based on the Chernoff-Hoeffding bound \cite{18}, we get
\begin{equation}
p(|I_{CHSH}-\bar{I}_{CHSH}|\geq\xi)\leq \exp(-2\xi^2\lceil\gamma N\rceil).
\end{equation}
Let $\varepsilon_{CHSH}=\exp(-2\xi^2\lceil\gamma N\rceil)$, we get
\begin{equation}
\xi=\sqrt{\frac{1}{2\lceil\gamma N\rceil}\ln \frac{1}{\varepsilon_{CHSH}}}.
\end{equation}

Further, denote $I'_{CHSH}=\frac{1}{\lfloor(1-\gamma) N\rfloor}\sum_{i\in\Gamma_{QPQ}}I_{i}$ and
$I=\frac{1}{ N}\sum_{i\in\{1,...,N\}}I_{i}$, we have
\begin{equation}
p(|I_{CHSH}-I|\geq v)\leq \exp(\frac{-2v^2\lceil\gamma N\rceil N}{N-\lceil\gamma N\rceil+1}),
\label{c1}
\end{equation}

Due to $I=\frac{1}{ N}\sum_{i\in\{1,...,N\}}I_{i}=\frac{\lceil\gamma N\rceil }{N}I_{CHSH}+\frac{\lfloor(1-\gamma) N\rfloor}{N}I'_{CHSH}$,
Eq. (\ref{c1}) can be rewritten as
\begin{equation}
\begin{split}
&p(|I_{CHSH}-I|\geq v)\\
&=p(|I_{CHSH}-\frac{\lceil\gamma N\rceil}{N}I_{CHSH}-\frac{\lfloor(1-\gamma) N\rfloor}{N}I'_{CHSH}|\geq v)\\
&=p(|\frac{\lfloor(1-\gamma) N\rfloor}{N}(I_{CHSH}-I'_{CHSH})|\geq v)\\
&\leq \exp(\frac{-2v^2\lceil\gamma N\rceil N}{N-\lceil\gamma N\rceil+1}).
\end{split}
\end{equation}

Then, we have
\begin{equation}
p(|I_{CHSH}-I'_{CHSH}|\geq \frac{N}{\lfloor(1-\gamma)N\rfloor} v)\leq \varepsilon_{QPQ},
\end{equation}
Let $\varepsilon_{QPQ}=\exp(\frac{-2v^2\lceil\gamma N\rceil N^3}{(N-\lceil\gamma N\rceil+1)\lfloor(1-\gamma)N\rfloor^2})$,
hence,
\begin{equation}
 v=\sqrt{\frac{(N-\lceil\gamma N\rceil+1)\lfloor(1-\gamma)N\rfloor^2
\ln\frac{1}{\varepsilon_{QPQ}}}{2\lceil\gamma N\rceil N^3}}.
\end{equation}
\qed

\begin{proposition}
For DI-QPQ protocol without the assumption of perfect detectors, under the condition that the relation $I_{CHSH}<\bar{I}_{CHSH}$ is violated for the subset $\Gamma_{CHSH}$, Bob can proceed
to the QPQ part for the remaining subset $\Gamma_{QPQ}$ securely when $N\rightarrow \infty$.
\end{proposition}
\label{pro2}
Note that $\Gamma_{CHSH} (\Gamma_{QPQ})$ represents the subset of entangled states that  performs the local CHSH Bell test(QPQ part),
$I_{CHSH}$ represents the observed value of CHSH Bell correlation function, and $\bar{I}_{CHSH}$ represents the  value of CHSH Bell correlation function in
predetermined states and measurements.

To sum up, on the basis of Theorem 1, when $N\rightarrow \infty$, the expression of $\xi$ tends to 0.
This indicates that the certified states in the $\Gamma_{CHSH}$ are in the predetermined form.
Furthermore,
the states in the $\Gamma_{QPQ}$
 can be certified, although the states in the $\Gamma_{QPQ}$ are not executed local CHSH Bell test.
Associated with proposition 1,  our  protocol is secure.

\section{Discussion and Conclusions}
\subsection{Discussion}
When  the state shared Alice and Bob
is $\epsilon$-close to the one given in Eq.(\ref{e1}), our protocol also works.

The state shared between them is descried as
\begin{equation}
\alpha|0\rangle_{B}|\phi_{0}\rangle_{A}+\beta|1\rangle_{B}|\phi_{1}\rangle_{A},
\label{e6}
\end{equation}
where
\begin{equation}
\begin{split}
&\alpha=\sqrt{\frac{1}{2}+\epsilon}, \qquad
\beta=\sqrt{\frac{1}{2}-\epsilon},\\
 &\epsilon\in (-\frac{1}{2}, 0)\cup (0, \frac{1}{2}),
 \end{split}
 \end{equation}

denoted as $\widetilde{\Psi}\rangle_{BA}$. When $\epsilon=0$, then $\widetilde{\Psi}\rangle_{BA}=\Psi\rangle_{BA}$.

If Alice and Bob agree to use $\widetilde{\Psi}\rangle_{BA}$ for performing
our protocol, our protocol is unchanging except for the value $\bar{I}_{CHSH}$.
the value $\bar{I}_{CHSH}$ is changed to
\begin{equation}
\begin{split}
\bar {I}_{CHSH}=&\frac{8-2[\sin\theta(\sin\psi_{1}+\sin\psi_{2})+2\alpha\beta}{\eta}\\
&\frac{(\cos\psi_{1}-\cos\psi_{2})]}{}+3[\sin\theta(\sin\psi_{1}+\sin\psi_{2})+\\
&2\alpha\beta(\cos\psi_{1}-\cos\psi_{2})]-8.
\end{split}
\end{equation}

When completing this protocol, Bob knows that the number of Alice's exacted keys is $(\frac{1}{2}+2\epsilon^2)\sin^2\theta|K|$. Thus, Bob utilizes a suitable postprocessing such that Alice's known key bits are reduced to 1 bit.
Obviously, Alice does not obtain more items than the item that she queried. Further, this protocol is secure.

It is not hard to find that $\widetilde{\Psi}\rangle_{BA}$ is the same as the state in the adversary scenario in III.A.
Compared with  the case in adversary scenario, the difference lies in:

\noindent 1)Alice and Bob agree to use $\widetilde{\Psi}\rangle_{BA}$ for carrying out DI-QPQ protocol;

\noindent 2)  Bob certifies whether the shared states are $\widetilde{\Psi}\rangle_{BA}$ or not;

\noindent 3) Different from Eq.(6) and Eq.(18), the critical value (i.e., $\bar{I}_{CHSH}$) is changed;

\noindent 4) Bob knows that the number of Alice's known keys is not $\frac{\sin^2\theta}{2}|K|$ but $(\frac{1}{2}+2\epsilon^2)\sin^2\theta|K|$,
which helps to choose appropriate postprocessing.
\subsection{Conlusion}
In this paper,  we analyzed the security of MRT17 protocol when the detectors were imperfect and
found this protocol was insecure in the above case.
Furthermore, we proposed DI-QPQ  protocol without  the assumption of perfect detectors.
Compared with MRT17 protocol, our  protocol is towards  practical and maintains the
security in the DI framework.

But,  some issues in DI-QPQ protocol are deserved to investigate in the future.

\noindent(1) Bell tests play an important role in the DI framework. While, Bell test encounters some loopholes such as free of choice loophole,detection loophole. We  investigate or construct loophole-free Bell test to be  adequate for DI-QPQ protocol.

\noindent(2) Novel DI-QPQ protocol which achieves better performance need be proposed. For example, a multi-bit block from the database in one query can be retrieved.

\section*{acknowledgement}
 We appreciate the anonymous reviewers for their valuable suggestions and  are grateful to Ya Cao, Runze Li for providing materials and helpful discussions.
This work is supported by NSFC (Grant Nos. 61802033,  61672110, 61702469, 61771439, 61701553), National Cryptography Development Fund (Grant No. MMJJ20170120), Sichuan Youth Science and Technology Foundation(Grant No. 2017JQ0045).

\section{Appendix}
\begin{center}
\textbf{Appendix A: The result of Eq. (\ref{e3}) cannot violate the relation (\ref{tt2})}
\label{A1}
\end{center}

First, we give the deduction of Eq. (\ref{e3}) as follows.

\begin{equation}
\begin{split}
&p(a_{i}\oplus b_{i}=x_{i}\wedge y_{i})\\
&=p_{X,Y}(0,0)[p(0,0|0,0)+p(1,1|0,0)]\\
&\quad+p_{X,Y}(0,1)[p(0,0|0,1)+p(1,1|0,1)]\\
&\quad+p_{X,Y}(1,0)[p(0,0|1,0)+p(1,1|1,0)]\\
&\quad+p_{X,Y}(1,1)[p(0,1|1,1)+p(1,0|1,1)]\\
&=\frac{1}{4}[\alpha^2 \cos^2(\frac{\theta-\psi_{1}}{2})+\beta^2 \sin^2(\frac{\theta+\psi_{1}}{2})]\\
&\quad+\frac{1}{4}[\alpha^2 \cos^2(\frac{\theta-\psi_{2}}{2})+\beta^2 \sin^2(\frac{\theta+\psi_{2}}{2})]\\
&\quad+\frac{1}{4}[(\frac{\alpha+\beta}{\sqrt{2}}\cos\frac{\psi_{1}}{2}\cos\frac{\theta}{2}
+\frac{\alpha-\beta}{\sqrt{2}}\sin\frac{\psi_{1}}{2}\sin\frac{\theta}{2})^2\\
&\quad+(\frac{\alpha-\beta}{\sqrt{2}}\sin\frac{\psi_{1}}{2}\cos\frac{\theta}{2}
-\frac{\alpha+\beta}{\sqrt{2}}\cos\frac{\psi_{1}}{2}\sin\frac{\theta}{2})^2]\\
&\quad+\frac{1}{4}[(\frac{\alpha+\beta}{\sqrt{2}}\sin\frac{\psi_{2}}{2}\cos\frac{\theta}{2}
-\frac{\alpha-\beta}{\sqrt{2}}\cos\frac{\psi_{2}}{2}\sin\frac{\theta}{2})^2\\
&\quad+(\frac{\alpha-\beta}{\sqrt{2}}\cos\frac{\psi_{2}}{2}\cos\frac{\theta}{2}
+\frac{\alpha+\beta}{\sqrt{2}}\sin\frac{\psi_{2}}{2}\sin\frac{\theta}{2})^2]\\
\end{split}
\end{equation}

\begin{equation}
\begin{split}
&=\frac{1}{8}\sin\theta(\sin\psi_{1}+\sin\psi_{2})+\frac{\sqrt{\frac{1}{4}-\epsilon^2}}{4}(\cos\psi_{1}-\cos\psi_{2})\\
&\quad+\frac{\cos\theta}{4}(\cos\psi_{1}+\cos\psi_{2})\epsilon+\frac{1}{2}\\
&=\frac{1}{8}\sin\theta(\sin\psi_{1}+\sin\psi_{2})+\frac{\sqrt{\frac{1}{4}-\epsilon^2}}{4}(\cos\psi_{1}-\cos\psi_{2})\\
&\quad+\frac{1}{2},
\end{split}
\end{equation}
where $p(0,0|0,0)=tr(|\Psi_{BA}\rangle\langle\Psi_{BA}||0\rangle\langle0|\otimes|\psi_{1}\rangle\langle\psi_{1}|)$,
other probabilities have similar operation method. The last equality holds since $\cos\psi_{1}+\cos\psi_{2}=0$.

Next, we prove that the above result cannot violate the relation (\ref{tt2}).
Denote the result of Eq. (\ref{e3}) as $f_{1}$, and let $\frac{1}{8}[\sin\theta(\sin\psi_{1}+\sin\psi_{2})+\cos\psi_{1}-\cos\psi_{2}]+\frac{1}{2}$ be $f_{0}$.
\begin{equation}
\begin{split}
F(\epsilon)&=f_{1}-f_{0}\\
&=\frac{2\sqrt{\frac{1}{4}-\epsilon^2}-1}{8}(\cos\psi_{1}-\cos\psi_{2}).
\end{split}
\end{equation}

Via derivation of $F$ with respect to $\epsilon$, we get
\begin{equation}
F'(\epsilon)=
-\frac{1}{4}(\cos\psi_{1}-\cos\psi_{2})\frac{\epsilon}{\sqrt{\frac{1}{4}-\epsilon^2}}.
\end{equation}
Case 1) when $\epsilon\in (0, \frac{1}{2})$, we get
 \begin{equation}
 \begin{split}
 &F'(\epsilon)<0,\\
 &F(0)=0.
 \end{split}
 \end{equation}

 \noindent Case 2) when $\epsilon\in (-\frac{1}{2}, 0)$, we get 
 \begin{equation}
 \begin{split}
 &F'(\epsilon)>0,\\
 &F(0)=0.
 \end{split}
 \end{equation}
Hence, when $\epsilon\in (-\frac{1}{2}, 0)\cup (0, \frac{1}{2})$, 
we get $F(\epsilon)<0$. That is, $f_{1}<f_{0}$.
Therefore, Eq. (\ref{e3}) cannot violate the relation Eq. (\ref{tt2}).
\qed
\begin{center}
\textbf{Appendix B: the deduction of Eq. (\ref{e5}) }
\end{center}

We deduce  $E(X_{0}Y_{0})$, $E(X_{0}Y_{1})$, $E(X_{1}Y_{0})$ and $E(X_{1}Y_{1})$, respectively.
\begin{equation}
\begin{split}
E(X_{0}Y_{0})&=\sum_{a,b\in\{0,1\}}(-1)^{ab}p(a,b|X_{0},Y_{0})\\
&=\sum_{a',b'\in\{1,-1\}}a'b'p(a',b'|X_{0},Y_{0})\\
&=p(1,1|X_{0},Y_{0})+p(-1,-1|X_{0},Y_{0})\\
&-p(-1,1|X_{0},Y_{0})-p(1,-1|X_{0},Y_{0})\\
&=\frac{1}{2}\cos^{2}(\frac{\theta-\psi_{1}}{2})
+\frac{1}{2}\sin^{2}(\frac{\theta+\psi_{1}}{2})\\
&-\frac{1}{2}\sin^{2}(\frac{\theta-\psi_{1}}{2})
-\frac{1}{2}\cos^{2}(\frac{\theta+\psi_{1}}{2})\\
&=\frac{1}{2}\cos(\theta-\psi_{1})-\frac{1}{2}\cos(\theta+\psi_{1})\\
&=\sin\theta\sin\psi_{1},
\label{1}
\end{split}
\end{equation}
where $p(1,1|X_{0},Y_{0})=tr(|\Psi_{BA}\rangle\langle\Psi_{BA}||0\rangle\langle0|\otimes|\psi_{1}\rangle\langle\psi_{1}|)$,
other probabilities have similar operation methods.
\begin{equation}
\begin{split}
E(X_{0}Y_{1})&=\sum_{a',b'\in\{1,-1\}}a'b'p(a',b'|X_{0},Y_{1})\\
&=p(1,1|X_{0},Y_{1})+p(-1,-1|X_{0},Y_{1})\\
&-p(-1,1|X_{0},Y_{1})-p(1,-1|X_{0},Y_{1})\\
&=\frac{1}{2}\cos^{2}(\frac{\theta-\psi_{2}}{2})
+\frac{1}{2}\sin^{2}(\frac{\theta+\psi_{2}}{2})\\
&-\frac{1}{2}\sin^{2}(\frac{\theta-\psi_{2}}{2})
-\frac{1}{2}\cos^{2}(\frac{\theta+\psi_{2}}{2})\\
&=\frac{1}{2}\cos(\theta-\psi_{2})-\frac{1}{2}\cos(\theta+\psi_{2})\\
&=\sin\theta\sin\psi_{2},
\label{2}
\end{split}
\end{equation}
where $p(1,1|X_{0},Y_{1})=tr(|\Psi_{BA}\rangle\langle\Psi_{BA}||0\rangle\langle0|\otimes|\psi_{2}\rangle\langle\psi_{2}|)$,
other probabilities have similar operation methods.
\begin{equation}
\begin{split}
E(X_{1}Y_{0})&=\sum_{a',b'\in\{1,-1\}}a'b'p(a',b'|X_{1},Y_{0})\\
&=p(1,1|X_{1},Y_{0})+p(-1,-1|X_{1},Y_{0})\\
&-p(-1,1|X_{1},Y_{0})-p(1,-1|X_{1},Y_{0})\\
&=\cos^{2}(\frac{\theta}{2})\cos^{2}(\frac{\psi_{1}}{2})
+\sin^{2}(\frac{\theta}{2})\cos^{2}(\frac{\psi_{1}}{2})\\
&-\cos^{2}(\frac{\theta}{2})\sin^{2}(\frac{\psi_{1}}{2})
-\sin^{2}(\frac{\theta}{2})\cos^{2}(\frac{\psi_{1}}{2})\\
&=\cos^2(\frac{\theta}{2})\cos(\psi_{1})+\sin^2(\frac{\theta}{2})\cos(\psi_{1})\\
&=\cos\psi_{1},
\label{3}
\end{split}
\end{equation}
where $p(1,1|X_{1},Y_{0})=tr(|\Psi_{BA}\rangle\langle\Psi_{BA}||+\rangle\langle+|\otimes|\psi_{1}\rangle\langle\psi_{1}|)$,
other probabilities have similar operation methods.
\begin{equation}
\begin{split}
E(X_{1}Y_{1})&=\sum_{a',b'\in\{1,-1\}}a'b'p(a',b'|X_{1},Y_{1})\\
&=p(1,1|X_{1},Y_{1})+p(-1,-1|X_{1},Y_{1})\\
&-p(-1,1|X_{1},Y_{1})-p(1,-1|X_{1},Y_{1})\\
&=\cos^{2}(\frac{\theta}{2})\cos^{2}(\frac{\psi_{2}}{2})
+\sin^{2}(\frac{\theta}{2})\cos^{2}(\frac{\psi_{2}}{2})\\
&-\cos^{2}(\frac{\theta}{2})\sin^{2}(\frac{\psi_{2}}{2})
-\sin^{2}(\frac{\theta}{2})\cos^{2}(\frac{\psi_{2}}{2})\\
&=\cos^2(\frac{\theta}{2})\cos(\psi_{1})+\sin^2(\frac{\theta}{2})\cos(\psi_{1})\\
&=\cos\psi_{2},
\label{4}
\end{split}
\end{equation}
where $p(1,1|X_{1},Y_{1})=tr(|\Psi_{BA}\rangle\langle\Psi_{BA}||+\rangle\langle+|\otimes|\psi_{2}\rangle\langle\psi_{2}|)$,
other probabilities have similar operation methods.

Based on the above the results (\ref{1})-(\ref{4}), we get
\begin{equation}
\begin{split}
\hat I_{CHSH}&=E(X_{0}Y_{0})+E(X_{0}Y_{1})+E(X_{1}Y_{0})-E(X_{1}Y_{1})\\
&=\sin\theta(\sin\psi_{1}+\sin\psi_{2})+\cos\psi_{1}-\cos\psi_{2}.
\end{split}
\end{equation}
\qed

\begin{center}
\textbf{Appendix C: Eq. (\ref{e4}) cannot violate the relation  (\ref{tt3}) }
\end{center}

The deduction of (\ref{e4}) is similar to that of (\ref{e5}).
Different from Eq. (\ref{e5}), the deduction of each probability term in Eq.(\ref{e4}) uses the state (\ref{e6}) instead of $|\Psi_{BA}\rangle$.
Thus, we get
\begin{equation}
\begin{split}
&E(X_{0}Y_{0})=\sin\theta\sin\psi_{1}+2\cos\theta\cos\psi_{1}\epsilon,\\
&E(X_{0}Y_{1})=\sin\theta\sin\psi_{2}+2\cos\theta\cos\psi_{2}\epsilon,\\
&E(X_{1}Y_{0})=2\sqrt{\frac{1}{4}-\epsilon^2}\cos\psi_{1},\\
&E(X_{1}Y_{1})=2\sqrt{\frac{1}{4}-\epsilon^2}\cos\psi_{2}.\\
\end{split}
\end{equation}
Hence, we have
\begin{equation}
\begin{split}
I_{CHSH}=&\sin\theta(\sin\psi_{1}+\sin\psi_{2})+2\cos\theta(\cos\psi_{1}+\cos\psi_{2})\epsilon\\
&+2\sqrt{\frac{1}{4}-\epsilon^2}(\cos\psi_{1}-\cos\psi_{2})\\
=&\sin\theta(\sin\psi_{1}+\sin\psi_{2})
+2\sqrt{\frac{1}{4}-\epsilon^2}(\cos\psi_{1}-\cos\psi_{2}),
\end{split}
\end{equation}
the last equality holds since $\cos\psi_{1}+\cos\psi_{2}=0$.

Next, we prove that the above result cannot violate the relation (\ref{e5}).
Denote the result of Eq. (\ref{e4}) as $g_{1}$, and let $\sin\theta(\sin\psi_{1}+\sin\psi_{2})+\cos\psi_{1}-\cos\psi_{2}$ be $g_{0}$.
\begin{equation}
\begin{split}
F(\epsilon)&=g_{1}-g_{0}\\
&=(2\sqrt{\frac{1}{4}-\epsilon^2}-1)(\cos\psi_{1}-\cos\psi_{2}).
\end{split}
\end{equation}

Via derivation of $F$ with respect to $\epsilon$, we get
\begin{equation}
F'(\epsilon)=
-2(\cos\psi_{1}-\cos\psi_{2})\frac{\epsilon}{\sqrt{\frac{1}{4}-\epsilon^2}}.
\end{equation}
Case 1) when $\epsilon\in (0, \frac{1}{2})$, we have 
 \begin{equation}
 \begin{split}
 &F'(\epsilon)<0,\\
 &F(0)=0.
 \end{split}
 \end{equation}

\noindent Case 2) when $\epsilon\in (-\frac{1}{2}, 0)$, we have 
 \begin{equation}
 \begin{split}
 &F'(\epsilon)>0,\\
 &F(0)=0.
 \end{split}
 \end{equation}
Hence, when $\epsilon\in (-\frac{1}{2}, 0)\cup (0, \frac{1}{2})$,
 we get $F(\epsilon)<0$. That is, $g_{1}<g_{0}$.
In short, Eq. (\ref{e4}) cannot violate the relation Eq. (\ref{e5}).
\qed

\begin{center}
\textbf{Appendix D: Eq. (\ref{e21}) violates the relation  (\ref{tt3}) }
\end{center}

First, we deduce  Eq. (\ref{e21}) in the following.

In the perfect detector scenario (i.e., $\eta=1$), when the state is in the form (\ref{e6}), Eq. (\ref{e4}) holds.
When $\eta\neq1$, Eq. (\ref{e4}) can be rewritten as
\begin{equation}
\begin{split}
&E(X_{0}Y_{0}|\Lambda_{0})+E(X_{0}Y_{1}|\Lambda_{0})+E(X_{1}Y_{0}|\Lambda_{0})-E(X_{1}Y_{1}|\Lambda_{0})\\
&=\sin\theta(\sin\psi_{1}+\sin\psi_{2})+2\sqrt{\frac{1}{4}-\epsilon^2}(\cos\psi_{1}-\cos\psi_{2}),
\end{split}
\label{s022}
\end{equation}
where $\Lambda_{0}=\Lambda_{X_{0}Y_{0}X_{1}Y_{1}}$ represents the ensemble that an arbitrary measurement pair gives the nonempty outcomes
(i.e., $a', b' \in\{+1, -1\}$).

$E(X_{j}Y_{k}|\Lambda_{0})$ is inaccessible in the actual experiment, while $E(X_{j}Y_{k}|\Lambda_{X_{j}Y_{k}})$ is calculated easily
in the above situation. So, in order  to get the relation between $E(X_{j}Y_{k}|\Lambda_{0})$ and $E(X_{j}Y_{k}|\Lambda_{X_{j}Y_{k}})$, we define
\begin{equation}
\delta=\min_{(j,k)}p(\Lambda_{0}|\Lambda_{X_{j}Y_{k}}),
\end{equation}
where $\min_{(j,k)}$ can be obtained by taking over all the measurement settings for $j,k\in\{0,1\}$.

The relation between $E(X_{j}Y_{k}|\Lambda_{X_{j}Y_{k}})$ and $E(X_{j}Y_{k}|\Lambda_{0})$ is given by
\begin{equation}
|E(X_{j}Y_{k}|\Lambda_{X_{j}Y_{k}})-\delta E(X_{j}Y_{k}|\Lambda_{0})|\leq 1-\delta
\label{e07}
\end{equation}
for $j, k\in\{0,1\}$.

\emph{proof}
Define $\overline{\Lambda_{0}}=\Lambda_{X_{j}Y_{k}}\setminus \Lambda_{0}$, which satisfies that
\begin{equation}
\begin{split}
&\Lambda_{0}\cap \overline{\Lambda_{0}}=\emptyset,\\
&\Lambda_{0}\cup \overline{\Lambda_{0}}=\Lambda_{X_{j}Y_{k}}.
\end{split}
\end{equation}
Hence, we can get
\begin{equation}
\begin{split}
&E(X_{j}Y_{k}|\Lambda_{X_{j}Y_{k}})
=E(X_{j}Y_{k}|\overline{\Lambda_{0}}\cup\Lambda_{0})\\
=&p(\overline{\Lambda_{0}}|\Lambda_{X_{j}Y_{k}})E(X_{j}Y_{k}|\overline{\Lambda_{0}})
+p(\Lambda_{0}|\Lambda_{X_{j}Y_{k}})E(X_{j}Y_{k}|\Lambda_{0}).
\end{split}
\end{equation}

Next, we have
\begin{equation}
\begin{split}
&|E(X_{j}Y_{k}|\Lambda_{X_{j}Y_{k}})-\delta E(X_{j}Y_{k}|\Lambda_{0})|\\
=& |p(\overline{\Lambda_{0}}|\Lambda_{X_{j}Y_{k}})E(X_{j}Y_{k}|\overline{\Lambda_{0}})
+p(\Lambda_{0}|\Lambda_{X_{j}Y_{k}})E(X_{j}Y_{k}|\Lambda_{0})\\
&-\delta E(X_{j}Y_{k}|\Lambda_{0})|\\
\leq& |p(\overline{\Lambda_{0}}|\Lambda_{X_{j}Y_{k}})E(X_{j}Y_{k}|\overline{\Lambda_{0}})|
+|p(\Lambda_{0}|\Lambda_{X_{j}Y_{k}})E(X_{j}Y_{k}|\Lambda_{0})\\
&-\delta E(X_{j}Y_{k}|\Lambda_{0})|\\
\leq& p(\overline{\Lambda_{0}}|\Lambda_{X_{j}Y_{k}})E(|X_{j}Y_{k}||\overline{\Lambda_{0}})
+(p(\Lambda_{0}|\Lambda_{X_{j}Y_{k}})-\delta)\\
&E(|X_{j}Y_{k}||\Lambda_{0})\\
\leq& 1-\delta,
\end{split}
\end{equation}
where $E(|X_{j}Y_{k}|)=\sum_{a', b'}|a'b'|p(a', b'|X_{j}, Y_{k})$.
 The first inequality holds based on $|x+y-z|\leq|x|+|y-z|$.
 The second inequality holds based on $|E(xy)|\leq E(|xy|)$.
The last inequality holds on the basis of $E(|X_{j}Y_{k}||\overline{\Lambda_{0}})=1$ and $E(|X_{j}Y_{k}||\Lambda_{0})=1$
since $|a'b'|=1$ for $a', b'\in \{+1,-1\}$.
\qed

Furthermore, in an actual experiment, we get
\begin{equation}
\begin{split}
&I_{CHSH}\\
=&|E(X_{0}Y_{0}|\Lambda_{X_{0}Y_{0}})+E(X_{0}Y_{1}|\Lambda_{X_{0}Y_{1}})+E(X_{1}Y_{0}|\Lambda_{X_{1}Y_{0}})\\
&-E(X_{1}Y_{1}|\Lambda_{X_{1}Y_{1}})|\\
=&|E(X_{0}Y_{0}|\Lambda_{X_{0}Y_{0}})-\delta E(X_{0}Y_{0}|\Lambda_{0})+\delta E(X_{0}Y_{0}|\Lambda_{0})\\
+&E(X_{0}Y_{1}|\Lambda_{X_{0}Y_{1}})-\delta E(X_{0}Y_{1}|\Lambda_{0})+\delta E(X_{0}Y_{1}|\Lambda_{0})\\
+&E(X_{1}Y_{0}|\Lambda_{X_{1}Y_{0}})-\delta E(X_{1}Y_{0}|\Lambda_{0})+\delta E(X_{1}Y_{0}|\Lambda_{0})\\
-&(E(X_{1}Y_{1}|\Lambda_{X_{1}Y_{1}})+\delta E(X_{1}Y_{1}|\Lambda_{0})-\delta E(X_{1}Y_{1}|\Lambda_{0}))|.
\end{split}
\label{e08}
\end{equation}

 Via the property of the absolute value,  Eq. (\ref{e08}) can be rewritten as
\begin{equation}
\begin{split}
I_{CHSH}\leq
&|E(X_{0}Y_{0}|\Lambda_{X_{0}Y_{0}})-\delta E(X_{0}Y_{0}|\Lambda_{0})|\\
&+|E(X_{0}Y_{1}|\Lambda_{X_{0}Y_{1}})-\delta E(X_{0}Y_{1}|\Lambda_{0})|\\
&+|E(X_{1}Y_{0}|\Lambda_{X_{1}Y_{0}})-\delta E(X_{1}Y_{0}|\Lambda_{0})|\\
&+|(E(X_{1}Y_{1}|\Lambda_{X_{1}Y_{1}})-\delta E(X_{1}Y_{1}|\Lambda_{0})|\\
&+\delta |E(X_{0}Y_{0}|\Lambda_{0})
+ E(X_{0}Y_{1}|\Lambda_{0})+ E(X_{1}Y_{0}|\Lambda_{0})\\
&-E(X_{1}Y_{1}|\Lambda_{0})|.\\
\end{split}
\label{a01}
\end{equation}
By using Eq. (\ref{s022}) and Eq. (\ref{e07}), Eq. (\ref{a01}) can be deduced as
\begin{equation}
\begin{split}
I_{CHSH}\leq&4(1-\delta)+[\sin\theta(\sin\psi_{1}+\sin\psi_{2})+2\sqrt{\frac{1}{4}-\epsilon^2}\\
&(\cos\psi_{1}-\cos\psi_{2})]\delta\\
=&4+[\sin\theta(\sin\psi_{1}+\sin\psi_{2})+2\sqrt{\frac{1}{4}-\epsilon^2}\\
&(\cos\psi_{1}-\cos\psi_{2})-4]\delta.
\end{split}
\label{e99}
\end{equation}

Next, we give the relation between $\delta$ and detection efficiency $\eta$ as follow:
\begin{equation}
\delta\geq 3-\frac{2}{\eta}.
\label{e100}
\end{equation}

\emph{proof}
For $j'\neq j$, $p(\Lambda_{X_{j'}}|\Lambda_{X_{j}Y_{k}})$ is given by
\begin{equation}
\begin{split}
&p(\Lambda_{X_{j'}}|\Lambda_{X_{j}Y_{k}})
=\frac{p(\Lambda_{X_{j'}Y_{j}}|\Lambda_{Y_{k}})}{p(\Lambda_{X_{j}}|\Lambda_{Y_{k}})}\\
=&\frac{p(\Lambda_{X_{j'}}|\Lambda_{B_{k}})+p(\Lambda_{X_{j}}|\Lambda_{Y_{k}})
-p(\Lambda_{X_{j'}}\cup\Lambda_{X_{j}}|\Lambda_{Y_{k}})}{p(\Lambda_{X_{j}}|\Lambda_{Y_{k}})}\\
\geq & \frac{2\eta-1}{\eta}.
\label{22}
\end{split}
\end{equation}
where  we assume that the detection efficiency of each party is independent and constant rate,
that is, $p(\Lambda_{X_{j'}}|\Lambda_{Y_{k}})=p(\Lambda_{X_{j}}|\Lambda_{Y_{k}})=\eta$.

Furthermore, when $j'\neq j$ and $k'\neq k$, $p(\Lambda_{X_{j'}Y_{k'}}|\Lambda_{X_{j}Y_{k}})$ can be deduced by
\begin{equation}
\begin{split}
p(\Lambda_{X_{j'}Y_{k'}}|\Lambda_{X_{j}Y_{k}})
&=p(\Lambda_{X_{j'}}|\Lambda_{X_{j}Y_{k}})+p(\Lambda_{Y_{k'}}|\Lambda_{X_{j}Y_{k}})\\
&-p(\Lambda_{X_{j'}}\cup\Lambda_{Y_{k'}}|\Lambda_{X_{j}Y_{k}})\\
&\geq \frac{2\eta-1}{\eta}+\frac{2\eta-1}{\eta}-1\\
 &= 3-\frac{2}{\eta}.
\end{split}
\label{232}
\end{equation}

 Finally, we calculate $\delta$.
Without loss of generality, assume that $j=1, k=1$, $p(\Lambda_{0}|\Lambda_{X_{1}Y_{1}})$ can be represented as
 \begin{equation}
\begin{split}
&p(\Lambda_{0}|\Lambda_{X_{1}Y_{1}})
=p(\Lambda_{X_{0}X_{1}Y_{0}Y_{1}}|\Lambda_{X_{1}Y_{1}})\\
=&p(\Lambda_{X_{0}Y_{0}}\cap\Lambda_{X_{1}Y_{1}}|\Lambda_{X_{1}Y_{1}})\\
=&p(\Lambda_{X_{0}Y_{0}}|\Lambda_{X_{1}Y_{1}})+p(\Lambda_{X_{1}Y_{1}}|\Lambda_{X_{1}Y_{1}})-p(\Lambda_{X_{0}Y_{0}}\cup\\
&\Lambda_{X_{1}Y_{1}}|\Lambda_{X_{1}Y_{1}})\\
\geq & p(\Lambda_{X_{0}Y_{0}}|\Lambda_{X_{1}Y_{1}})\\
 \geq & 3-\frac{2}{\eta},
\end{split}
\end{equation}
where the fifth line holds since $p(\Lambda_{X_{1}Y_{1}}|\Lambda_{X_{1}Y_{1}})=1,
p(\Lambda_{X_{0}Y_{0}}\cup\Lambda_{X_{1}Y_{1}}|\Lambda_{X_{1}Y_{1}})\leq 1.$
The sixth line holds based on Eq. (\ref{232}).

So, we get
\begin{equation}
\delta\geq 3-\frac{2}{\eta}.
\label{b1}
\end{equation}
\qed

Input Eq. (\ref{e100}) into Eq. (\ref{e99}), we obtain
\begin{equation}
\begin{split}
 I_{CHSH}&\leq \frac{8-2[\sin\theta(\sin\psi_{1}+\sin\psi_{2})+2\sqrt{\frac{1}{4}-\epsilon^2}}{\eta}\\
&\quad\frac{(\cos\psi_{1}-\cos\psi_{2})]}{}-8+3[\sin\theta(\sin\psi_{1}+\sin\psi_{2})\\
&+2\sqrt{\frac{1}{4}-\epsilon^2}(\cos\psi_{1}-\cos\psi_{2})].
\end{split}
\end{equation}
Here, in order to demonstrate there exists an attack strategy, we only consider $I_{CHSH}
=\frac{8-2[\sin\theta(\sin\psi_{1}+\sin\psi_{2})+2\sqrt{\frac{1}{4}-\epsilon^2}(\cos\psi_{1}-\cos\psi_{2})]}{\eta}
-8+3[\sin\theta(\sin\psi_{1}+\sin\psi_{2})+2\sqrt{\frac{1}{4}-\epsilon^2}(\cos\psi_{1}-\cos\psi_{2})]$.

Next, we prove that the above value violates the relation (\ref{tt3}) in some cases.
Let
\begin{align}
&A=\sin\theta(\sin\psi_{1}+\sin\psi_{2}),\\
&B=\cos\psi_{1}-\cos\psi_{2},\\
&C=(8-2A+B)\eta-8+2A.
\end{align}

On the basis of \cite{166}, the detection loophole of CHSH Bell test can be closed when $\eta>\frac{2}{1+\sqrt{2}}$.

(1) when $\eta<\frac{8-2A}{8-2A+B}$, we get
\begin{equation}
(8-2A+B)\eta-8+2A<0.
\end{equation}
 Now, when $\epsilon\in(-\frac{1}{2}, 0)\cup (0, \frac{1}{2})$, we get
 \begin{equation}
 2(3\eta-2)B\sqrt{\frac{1}{4}-\epsilon^2}>(8-2A+B)\eta-8+2A,
 \label{a11}
 \end{equation}
since $(3\eta-2)B\sqrt{\frac{1}{4}-\epsilon^2}>0$ and $(8-2A+B)\eta-8+2A<0$.

 In the above case, we get 
\begin{equation}
\begin{split}
Eq.(\ref{e21})-Eq.(\ref{e5})&=\frac{2(3\eta-2)B\sqrt{\frac{1}{4}-\epsilon^2}-[(8-2A}{\eta}\\
&\frac{\quad+B)\eta+8-2A]}{}.
\end{split}
\end{equation}

On the basis of the relation (\ref{a11}), we know
\begin{equation}
Eq.(\ref{e21})-Eq.(\ref{e5})>0.
\end{equation}
Hence, Eq. (\ref{e21}) can violate the relation Eq. (\ref{e5}) in the above case.

 (2) when $\frac{8-2A}{8-2A+B}<\eta<\frac{8-2A-2B}{8-2A-2B}$, we get
 \begin{align}
 (8-2A-B)\eta-8+2A>0,\\
(8-2A-2B)\eta<8-2A-2B.
\label{aa1}
\end{align}
Eq. (\ref{aa1}) can be rewritten as
$(3\eta-2)B-[(8-2A+B)\eta-8+2A]>0.$ That is, $(3\eta-2)B>C>0$.

Further, when $\epsilon \in (-\frac{\sqrt{(3\eta-2)^2B^2-C^2}}{2(3\eta-2)B}, 0)\cup (0, \frac{\sqrt{(3\eta-2)^2B^2-C^2}}{2(3\eta-2)B})$,
we get
\begin{equation}
4(3\eta-2)^2B^2\epsilon^2<(3\eta-2)^2B^2-C^2.
\end{equation}
So, we have
\begin{equation}
2(3\eta-2)B\sqrt{\frac{1}{4}-\epsilon^2}>C>0.
\label{a12}
\end{equation}
On the basis of the relation (\ref{a12}), we know
\begin{equation}
Eq. (\ref{e21})-Eq. (\ref{e5})>0.
\end{equation}
Hence, Eq. (\ref{e21}) can violate the relation Eq. (\ref{tt3}) in the above case.
\qed
\begin{center}
\textbf{Appendix E: The deduction of Eq.(\ref{s33})}
\end{center}
In the perfect detector scenario ($\eta=1$), we can assert that Eq. (\ref{e5}) has self-tested the state and measurements
in the predetermined forms. When $\eta$ is no larger than a certain value, Alice takes advantage of the detector inefficiency
such that Bob still gets the result (\ref{e5}). In fact, the state and measurements are not in the predetermined form.
 Now,
 Alice can get more information about items without being  discovered by Bob. In order to avoid this cheat strategy,
 more precisely, Eq. (\ref{e5}) can be rewritten as
\begin{equation}
\begin{split}
&E(X_{0}Y_{0}|\Lambda_{0})+E(X_{0}Y_{1}|\Lambda_{0})+E(X_{1}Y_{0}|\Lambda_{0})-E(X_{1}Y_{1}|\Lambda_{0})\\
&=\sin\theta(\sin\psi_{1}+\sin\psi_{2})+\cos\psi_{1}-\cos\psi_{2},
\end{split}
\label{s22}
\end{equation}
where $\Lambda_{0}=\Lambda_{X_{0}Y_{0}X_{1}Y_{1}}$ represents the ensemble that an arbitrary measurement pair gives the nonempty outcomes
(i.e., $a',b'\in\{+1, -1\}$).

Next, the deduction of Eq. (\ref{s33}) is similar to that of Eq. (\ref{e21}). While, the difference between them lies in the relation (\ref{e99}).
In the deduction of Eq. (\ref{s33}), $I_{CHSH}$ can be rewritten as
\begin{equation}
\begin{split}
 I_{CHSH}\leq&4(1-\delta)+[\sin\theta(\sin\psi_{1}+\sin\psi_{2})+\cos\psi_{1}-\cos\psi_{2}]\delta\\
=&4+[\sin\theta(\sin\psi_{1}+\sin\psi_{2})+\cos\psi_{1}-\cos\psi_{2}-4]\delta.
\end{split}
\label{e9}
\end{equation}

Input Eq. (\ref{e100}) into Eq. (\ref{e9}), we obtain
\begin{equation}
\begin{split}
 I_{CHSH}&\leq\frac{8-2[\sin\theta(\sin\psi_{1}+\sin\psi_{2})+\cos\psi_{1}-\cos\psi_{2}]}{\eta}\\
&-8+3[\sin\theta(\sin\psi_{1}+\sin\psi_{2})+\cos\psi_{1}-\cos\psi_{2}].
\end{split}
\end{equation}
 Hence, denote $\bar I_{CHSH}$ as
$\frac{8-2[\sin\theta(\sin\psi_{1}+\sin\psi_{2})+\cos\psi_{1}-\cos\psi_{2}]}{\eta}$ $
-8+3[\sin\theta(\sin\psi_{1}+\sin\psi_{2})+\cos\psi_{1}-\cos\psi_{2}]$.

\qed


\end{document}